\documentstyle[aps,prl]{revtex} 
\draft  
 
\begin{document} 
\author{Xiang-Song Chen and Fan Wang} 
\address{Department of Physics and Center for Theoretical Physics,  
Nanjing University, Nanjing 210093, China} 
\title{Gauge Invariance and Hadron Structure} 
\maketitle 
 
\begin{abstract} 
We prove that the {\em gauge dependent} gluon spin, gluon and quark orbital 
angular momenta operators have {\em gauge invariant} expectation values on 
hadron states with {\em definite} momentum and polarization, therefore the 
conventional decomposition of nucleon spin into contributions from the spin 
and orbital angular momentum of quark and gluon is {\em gauge independent}. 
Similar conclusions apply to the {\em gauge dependent} quark momentum and 
kinetic energy operators, and accordingly nucleon momentum and mass 
structures. 
\end{abstract} 
 
 
Gauge invariance is considered as one of the most fundamental principles in 
modern physics. All physical laws and observables are believed to be 
invariant under gauge transformations, and hence should be given gauge 
invariant theoretical descriptions. Yet we sometimes do encounter gauge 
non-invariance. A case in point, which is well known in the contemporary 
extensive studies of nucleon spin structure, is that the gluon spin, gluon 
and quark orbital angular momenta operators are not separately gauge 
invariant. And in retrospect, the electron angular momentum operator whose 
expectation values are usually used to label the atomic energy levels, is 
not gauge invariant either. Then regarding the principle of gauge 
invariance, how should we, if not impossible, give definite physical 
meanings to these operators, and especially to the labeling of atomic energy 
levels which we have talked abut for many decades? 
 
One possible approach to solve this problem explored in the studies of 
nucleon spin structure is to find a particular gauge in which the gauge 
dependent operator can be related to experimental quantities. For example, 
in light-cone coordinates and light-cone gauge $A^{+}=0$ the gluon spin can 
be related to the polarized gluon distribution measured in hard QCD 
processes \cite{Balitsky,Jaffe}. In this paper, we will examine a more 
appealing possibility: since what are related to experimental quantities are 
not the quantum operators themselves, but their matrix elements on physical 
states, we will not worry about the gauge dependence of operators if they 
give gauge invariant matrix elements on the studied states. We find that 
this is indeed the case for the operators concerning hadron mass, momentum, 
and spin structures. These operators, even though gauge dependent, have 
gauge invariant expectation values on hadron states with definite momentum 
and polarization. 
 
To be specific, let's begin with the nucleon spin structure. The QCD 
rotation generator can be separated into four parts  
\begin{eqnarray} 
{\vec J}_{QCD} &=&\frac 12\int d^3x\psi ^{\dagger }{\vec \Sigma }\psi +\int 
d^3x\psi ^{\dagger }{\vec x}\times \frac 1i{\vec \partial }\psi  \nonumber \\ 
&& +2\int d^3xTr\left\{ {\vec E}\times {\vec A}\right\}  
+2\int d^3xTr\left\{ E_i{\vec x}\times {\vec \partial }A_i\right\}  
\nonumber \\ 
&\equiv &{\vec S}_q+{\vec L}_q+{\vec S}_g+{\vec L}_g,  \label{decomp1} 
\end{eqnarray} 
where and below any quantity is defined in the same way as in ref. \cite 
{Lee}, and repeated indices are summed over. They have the obvious 
interpretations of quark spin, quark orbital angular momentum, gluon spin, 
and gluon orbital angular momentum respectively in the canonical formalism. 
However, there exists the well known difficulty that ${\vec L}_q$, ${\vec S}%
_g$, and ${\vec L}_g$ are not separately gauge invariant, therefore their 
physical significance seems obscure. Great efforts have been devoted to 
overcoming this difficulty. One example is what we have mentioned above for 
gluon spin, to study the experimental significance of these operators in a 
particular gauge. And on the other hand, one may naturally want to find 
gauge invariant operators for gluon spin, gluon and quark orbital angular 
momenta. Ji \cite{Ji:spin} and ourselves \cite{Chen} have indeed obtained an 
explicitly gauge invariant separation of QCD angular momentum operator  
\begin{eqnarray} 
{\vec J}_{QCD} &=&\frac 12\int d^3x\psi ^{\dagger }{\vec \Sigma }\psi +\int 
d^3x\psi ^{\dagger }{\vec x}\times \frac 1i{\vec D}\psi  \nonumber \\ 
&& +2\int d^3xTr\left\{ {\vec x}\times \left( {\vec E}\times {\vec B}\right) 
\right\}  \nonumber \\ 
&\equiv &{\vec S}_q+{\vec L}_q^{\prime }+{\vec J}_g^{\prime }. 
\label{decomp2} 
\end{eqnarray} 
However, a further decomposition of ${\vec J}_g^{\prime }$ into gauge 
invariant gluon spin and orbital parts is lacking, and ${\vec L}_q^{\prime }$ 
and ${\vec J}_g^{\prime }$ do not obey the ${\vec J}\times {\vec J}=i{\vec J} 
$ algebra even when $\psi $ and $A$ refer to bare fields, although ${\vec J}%
_g^{\prime }$ does in a pure gauge field theory. Therefore one should even 
hesitate a little to call them angular momentum operators. 
 
Now go back to the gauge dependence of the decomposition in eq.(\ref{decomp1}%
). Does it really mean that we are unable to discuss gluon spin, gluon 
and quark orbital angular momenta without referring to a particular gauge? 
And consequently, only in a particular gauge does the decomposition of 
nucleon spin into corresponding contributions make sense? Or actually makes 
no sense at all since gauge invariance is committed so much priority in a 
gauge theory like QCD? A more realistic problem: have we made great 
negligence in the scores of years of talking about quark or electron angular 
momentum in quark models or atomic physics? And they actually also have no 
definite physical meanings? 
 
We find that the problem is, fortunately, not so serious. The point is that 
although the gluon spin, gluon and quark orbital angular momenta, or 
electron angular momentum operators are all gauge dependent, their 
contributions to nucleon or atomic spin, which are defined as their 
expectation values on nucleon or atomic states with {\em definite} momentum 
and polarization, are all gauge ${\em independent}$. We give below the proof. 
 
To obtain the matrix element like $\left\langle p_1s_1\left| {\vec L}%
_q\right| p_2s_2\right\rangle $ (with $\left| ps\right\rangle $ a nucleon 
state with momentum $p$ and polarization $s$), we consider the following 
three-point Green function:  
\begin{equation} 
G_{N_l{\vec L}_qN_{l^{\prime }}^{\dagger }}^{p_1p_2}\equiv \int 
d^4x_1d^4x_2e^{-ip_1\cdot x_1}e^{-ip_2\cdot x_2}\left\langle 0\left| 
T\left\{ N_l(x_1){\vec L}_qN_{l'}^{\dagger }(x_2)\right\} \right| 
0\right\rangle , 
\label{G:def} 
\end{equation} 
where $N_l(x)$ is the nucleon interpolating operator. 
 
As $p_1$, $p_2$ go to the nucleon mass shell, $G_{N_l{\vec L}_qN_{l^{\prime 
}}^{\dagger }}^{p_1p_2}$ becomes  
\begin{eqnarray} 
G_{N_l{\vec L}_qN_{l^{\prime }}^{\dagger }}^{p_1p_2}\stackrel{%
p_{1,2}^2\rightarrow -m^2}{\longrightarrow }&&\frac{-2i\sqrt{\vec p_1^2+m^2}}{%
p_1^2+m^2-i\varepsilon }(2\pi )^3\sum_{ss^{\prime }}\left\langle 0\left| 
N_l(0)\right| p_1s\right\rangle \left\langle p_1s\left| {\vec L}_q\right| 
p_2s^{\prime }\right\rangle \nonumber \\
&&\times \left\langle p_2s^{\prime }\left| N_{l^{\prime 
}}^{\dagger }(0)\right| 0\right\rangle (2\pi )^3\frac{-2i\sqrt{\vec p_2^2+m^2%
}}{p_2^2+m^2-i\varepsilon }.  \label{G-lLl} 
\end{eqnarray} 
Using Lorentz invariance to write  
\begin{eqnarray} 
\left\langle 0\left| N_l(0)\right| p_1s\right\rangle &=&(2\pi 
)^{-3/2}Zu_l(p_1s),  \nonumber  \\ 
\left\langle p_2s^{\prime }\left| N_{l^{\prime }}^{\dagger }(0)\right| 
0\right\rangle &=&(2\pi )^{-3/2}Z^{*}u_{l^{\prime }}^{*}(p_2s^{\prime }), 
\label{Z:def} 
\end{eqnarray} 
where $u_l(ps)$ is the Dirac spinor with normalization 
$\sum_lu_l^{*}(ps)u_l(ps^{%
\prime })=\delta _{ss^{\prime }}$, and multiplying $G_{N_l{\vec L}%
_qN_{l^{\prime }}^{\dagger }}^{p_1p_2}$ with $(2\pi 
)^{-3/2}u_l^{*}(p_1s_1)(2\pi )^{-3/2}u_{l^{\prime }}(p_2s_2)$ and 
summing over $ll^{\prime }$, we get  
\begin{eqnarray} 
G_{N{\vec L}_qN^{\dagger }}^{p_1s_1p_2s_2} &&\equiv \sum_{ll^{\prime }}(2\pi 
)^{-3/2}u_l^{*}(p_1s_1)G_{N_l{\vec L}_qN_{l^{\prime }}^{\dagger 
}}^{p_1p_2}(2\pi )^{-3/2}u_{l^{\prime }}(p_2s_2)  \nonumber \\ 
&&\stackrel{p_{1,2}^2\rightarrow -m^2}{\longrightarrow }\frac{-2i\sqrt{\vec p%
_1^2+m^2}}{p_1^2+m^2-i\varepsilon }\left| Z\right| ^2\left\langle 
p_1s_1\left| {\vec L}_q\right| p_2s_2\right\rangle \frac{-2i\sqrt{\vec p%
_2^2+m^2}}{p_2^2+m^2-i\varepsilon }.  \label{G:NLN} 
\end{eqnarray} 
 
The matrix element in eq.(\ref{G:NLN}) is just what we want. The 
normalization factor $\left| Z\right| ^2$ can be obtained from the two-point 
Green function  
\begin{eqnarray} 
G_{N_lN_{l^{\prime }}^{\dagger }}^{p_1p_2} &&\equiv \int 
d^4x_1d^4x_2e^{-ip_1\cdot x_1}e^{-ip_2\cdot x_2}\left\langle 0\left| 
T\left\{ N_l(x_1)N_{l'}^{\dagger }(x_2)\right\} \right| 0\right\rangle  
\nonumber \\
&&\stackrel{p_1^2\rightarrow -m^2}{\longrightarrow }\frac{-2i\sqrt{\vec p%
_1^2+m^2}}{p_1^2+m^2-i\varepsilon }\left| Z\right| 
^2\sum_su_l(p_1s)u_{l^{\prime }}^{*}(p_1s)(2\pi )^4\delta ^4(p_1+p_2). 
\label{G:NN} 
\end{eqnarray} 
 
Eqs.(\ref{G:NLN}, \ref{G:NN}) show clearly that the gauge dependence of $%
\left\langle p_1s_1\left| {\vec L}_q\right| p_2s_2\right\rangle $ is 
equivalent to that of the ratio $G_{N{\vec L}_qN^{\dagger 
}}^{p_1s_1p_2s_2}/G_{N_lN_{l^{\prime }}^{\dagger }}^{p_1p_2}$. To explore 
the gauge dependence of $G_{N{\vec L}_qN^{\dagger }}^{p_1s_1p_2s_2}$ and $%
G_{N_lN_{l^{\prime }}^{\dagger }}^{p_1p_2}$, we use the method of functional 
integral to write  
\begin{eqnarray} 
G_{N{\vec L}_qN^{\dagger }}^{p_1s_1p_2s_2} &=&\int d^4x_1d^4x_2e^{-ip_1\cdot 
x_1}e^{-ip_2\cdot x_2}\sum_{ll^{\prime }}(2\pi )^{-3/2}u_l^{*}(p_1s_1)(2\pi 
)^{-3/2}u_{l^{\prime }}(p_2s_2)  \nonumber  \\ 
&&\times \int \prod_{n,x}d\phi _n(x)e^{iI}N_l(x_1){\vec L}_qN_{l'}^{\dagger 
}(x_2)B\left[ f[\phi ]\right] {\rm Det}{\cal F}\left[ \phi \right]  
\label{G:int} 
\end{eqnarray} 
(and likewise for $G_{N_lN_{l^{\prime }}^{\dagger }}^{p_1p_2}$),
where $\phi _n(x)$ denotes collectively the quark and gluon fields, $B\left[ 
f[\phi ]\right] $ is some functional of a general gauge-fixing functional $%
f_\alpha [\phi ;x]$, and ${\cal F}$ is defined as  
\begin{equation} 
{\cal F}_{\alpha x,\beta y}[\phi ]\equiv \frac{\delta f_\alpha [\phi 
_\lambda ;x]}{\delta \lambda ^{\beta }(y)}|_{\lambda =0},  \label{F:def} 
\end{equation} 
where $\phi _\lambda (x)$ is the result of $\phi (x)$ after a gauge 
transformation with parameters $\lambda ^{\alpha} (x)$. 
 
By saying that a matrix element or Green function is gauge independent, 
we mean that it is 
independent of the choice of the gauge-fixing functional $f_\alpha [\phi ;x]$ 
and the functional $B[f]$ up to an irrelevant field-independent constant 
factor. To demonstrate the gauge independence of $G_{N{\vec L}_qN^{\dagger 
}}^{p_1s_1p_2s_2}$ and $G_{N_lN_{l^{\prime }}^{\dagger }}^{p_1p_2}$, we 
first give a brief review of the standard proof of gauge independence for 
the Green function of {\em gauge invariant} 
operators $O_1O_2\cdots O_i$ 
\cite{Weinberg}:  
\begin{eqnarray} 
G_{O_1\cdots O_i}&\equiv &
\left\langle 0\left|
T\left\{ O_1\cdots O_i\right\} \right|
0\right\rangle \nonumber \\
&=& \int \prod_{n,x}d\phi _n(x)e^{iI}O_1 
\cdots O_iB\left[ f[\phi ]\right] {\rm Det}{\cal F}\left[ \phi \right] . 
\label{G:O} 
\end{eqnarray} 
 
Change the integration variable from $\phi $ to $\phi _\omega $, with $\omega 
^\alpha (x)$ an arbitrary set of gauge transformation parameters, and make 
use of the gauge invariance of the measure $\prod_{n,x}d\phi _n(x)$, the 
action $I$, and the operators $O_1\cdots O_i$, we have  
\begin{equation} 
G_{O_1\cdots O_i}=\int \prod_{n,x}d\phi _n(x)e^{iI}O_1 
\cdots O_iB\left[ f[\phi _\omega ]\right] {\rm Det}{\cal F}\left[ \phi 
_\omega \right] .  \label{G:O'} 
\end{equation} 
It can be verified that \cite{Weinberg}  
\begin{equation} 
{\rm Det}{\cal F}\left[ \phi _\omega \right] ={\rm Det}{\cal J}\left[ \phi 
,\omega \right] \cdot {\rm Det}{\cal R}\left[ \omega \right] ,  \label{Det} 
\end{equation} 
where  
\begin{equation} 
{\cal R}_{~~~\beta y}^{\alpha x}\left[ \omega \right] =\frac{\delta  
\widetilde{\omega }^\alpha (x;\omega ,\lambda )}{\delta \lambda ^\beta (y)}%
|_{\lambda =0}  \label{R} 
\end{equation} 
(with $\widetilde{\omega }^\alpha $ the parameters of the associate 
transformation $e^{-i\lambda (x)}e^{-i\omega (x)}\equiv e^{-i\widetilde{%
\omega }(x)}$) is solely a functional of $\omega ^\alpha (x)$, and  
\begin{equation} 
{\rm Det}{\cal J}\left[ \phi ,\omega \right] ={\rm Det}\frac{\delta f_\alpha 
[\phi _\omega ;x]}{\delta \omega ^\beta (y)}  \label{J} 
\end{equation} 
is just the Jacobian when transforming the integration variables from $%
\omega ^\beta (y)$ to $f_\alpha [\phi _\omega ;x]$. Therefore if we multiply 
both sides of eq.(\ref{G:O'}) with the weight functional  
\begin{equation} 
\rho [\omega ]=1/{\rm Det}{\cal R}\left[ \omega \right]   \label{rho} 
\end{equation} 
and integrate over $\omega ^\alpha (x)$, then by noting that $G_{O_1 
\cdots O_i}$ is independent of $\omega ^\alpha (x)$, we get  
\begin{equation} 
G_{O_1\cdots O_i}=\frac{\int \prod_{n,x}d\phi _n(x)e^{iI}O_1
\cdots O_iC[\phi ]}{\int \prod_{\alpha ,x}d\omega ^\alpha (x)\rho 
[\omega ]},  \label{G:final} 
\end{equation} 
where  
\begin{equation} 
C[\phi ]\equiv \int \prod_{\alpha ,x}d\omega ^\alpha (x)\rho [\omega 
]B\left[ f[\phi _\omega ]\right] {\rm Det}{\cal F}\left[ \phi _\omega 
\right]   \label{C:def} 
\end{equation} 
carries all the dependence of $G_{O_1 \cdots O_i}$ on the 
gauge-fixing functional $f[\phi ;x]$ and the functional $B\left[ f\right] $. 
Now make use of eqs.(\ref{Det}, \ref{rho}), we have  
\begin{eqnarray} 
C[\phi ] &=&\int \prod_{\alpha ,x}d\omega ^\alpha (x){\rm Det}{\cal J}\left[ 
\phi ,\omega \right] B\left[ f[\phi _\omega ]\right]   \nonumber \\ 
\  &=&\int \prod_{\alpha ,x}df_\alpha (x)B\left[ f\right] \equiv C. 
\label{C} 
\end{eqnarray} 
 
This is clearly independent of the choice of $f[\phi ,x]$, which has been 
reduced to a mere variable of integration, and it depends on $B\left[ 
f\right] $ only through a $\phi $-independent constant $C$. Thus we proved 
that $G_{O_1 \cdots O_i}$ is gauge independent as the operators $%
O_1\cdots O_i$ are gauge invariant. 
 
Now we come to $G_{N{\vec L}_qN^{\dagger }}^{p_1s_1p_2s_2}$ and $%
G_{N_lN_{l^{\prime }}^{\dagger }}^{p_1p_2}$. Since nucleon is color singlet, 
the nucleon interpolating operator $N_l(x)$ is SU(3) gauge invariant. 
Therefore according to the above proof, $G_{N_lN_{l^{\prime }}^{\dagger 
}}^{p_1p_2}$ is SU(3) gauge independent. However, ${\vec L}_q$ is SU(3) 
non-invariant. After an arbitrary gauge transformation  
\begin{eqnarray} 
\psi (x) &\rightarrow &U(x)\psi ,  \nonumber \\ 
A_\mu (x) &\rightarrow &U(x)A_\mu U^{-1}(x)+\frac igU(x)\partial _\mu 
U^{-1}(x)  \label{transform} 
\end{eqnarray} 
${\vec L}_q$ induces an extra term  
\begin{equation} 
\delta {\vec L}_q=\int d^3x\psi ^{\dagger }\left( U^{-1}{\vec x}\times \frac  
1i{\vec \partial }U\right) \psi ,  \label{deltaLq} 
\end{equation} 
where $U(x)=e^{-i\omega (x)}=e^{-i\omega ^\alpha (x)T^\alpha }$. So if we 
also change the integration variable from $\phi $ to $\phi _\omega $ as we 
did for $G_{O_1\cdots O_i}$, $G_{N{\vec L}_qN^{\dagger 
}}^{p_1s_1p_2s_2}$ becomes:  
\begin{eqnarray} 
G_{N{\vec L}_qN^{\dagger }}^{p_1s_1p_2s_2} &=&\int d^4x_1d^4x_2e^{-ip_1\cdot 
x_1}e^{-ip_2\cdot x_2}\sum_{ll^{\prime }}(2\pi )^{-3/2}u_l^{*}(p_1s_1)(2\pi 
)^{-3/2}u_{l^{\prime }}(p_2s_2)  \nonumber \\  
&&\times \int \prod_{n,x}d\phi _n(x)e^{iI}N_l(x_1)({\vec L}_q+\delta {\vec L}%
_q)N_{l'}^{\dagger }(x_2)B\left[ f[\phi _\omega ]\right] {\rm Det}{\cal F}
\left[ 
\phi _\omega \right]  \nonumber  \\ 
&\equiv &G_{{\vec L}_q}+G_{\delta {\vec L}_q}.  \label{G12} 
\end{eqnarray} 
 
Due to the extra term $G_{\delta {\vec L}_q}$, the above proof of gauge 
invariance does not simply apply. However, if we can show that $G_{\delta {%
\vec L}_q}$actually equals zero, then the above proof {\em applies} and 
hence $G_{N{\vec L}_qN^{\dagger }}^{p_1s_1p_2s_2}$ is gauge independent. 
 
Comparing the expression of $G_{\delta {\vec L}_q}$ with eqs.(\ref{G:int},  
\ref{G:NLN}, \ref{G:def}) shows  
\begin{eqnarray} 
G_{\delta {\vec L}_q} &&=\int d^4x_1d^4x_2e^{-ip_1\cdot x_1}e^{-ip_2\cdot 
x_2}\sum_{ll^{\prime }}(2\pi )^{-3/2}u_l^{*}(p_1s_1)(2\pi 
)^{-3/2}u_{l^{\prime }}(p_2s_2)\left\langle 0\left| T\left\{ N_l(x_1)\delta {%
\vec L}_q^{\prime }N_{l'}^{\dagger }(x_2)\right\} \right| 0\right\rangle  
\nonumber \\ 
&& \stackrel{p_{1,2}^2\rightarrow -m^2}{\longrightarrow }\frac{-2i\sqrt{\vec  
p_1^2+m^2}}{p_1^2+m^2-i\varepsilon }\left| Z\right| ^2\left\langle 
p_1s_1\left| \delta {\vec L}_q^{\prime }\right| p_2s_2\right\rangle \frac{-2i%
\sqrt{\vec p_2^2+m^2}}{p_2^2+m^2-i\varepsilon },  \label{G2} 
\end{eqnarray} 
where  
\begin{eqnarray} 
\delta {\vec L}_q^{\prime } &=&\int d^3x\left( U^{-1}\psi \right) ^{\dagger 
}\left( U^{-1}{\vec x}\times \frac 1i{\vec \partial }U\right) \left( 
U^{-1}\psi \right)  \nonumber \\ 
&=&-\int d^3x\psi ^{\dagger }\left( U{\vec x}\times \frac 1i{\vec \partial  
}U^{-1}\right) \psi .  \label{deltaLq'} 
\end{eqnarray} 
 
Without losing generality, we let the nucleon polarize along the third 
direction: $J_{QCD}^3\left| ps\right\rangle =s\left| ps\right\rangle $. We 
will demonstrate that $\delta L_q^{3\prime }$ can be expressed as a 
commutator with $J_{QCD}^3$, then eq.(\ref{G2}) shows clearly the result we 
aimed at: $G_{\delta {\vec L}_q}$ vanishes if $s_1=s_2$ and hence $G_{N{\vec  
L}_qN^{\dagger }}^{p_1sp_2s}$ is gauge independent. 
 
To obtain the commutator expression of $\delta {\vec L}_q^{\prime }$, we 
express the pure gauge form $U(x)\partial _\mu U^{-1}(x)$ in terms $\omega 
(x)$ \cite{Huang}  
\begin{equation} 
U\partial _\mu U^{-1}=e^{-i\omega }\partial _\mu e^{i\omega }=i\int_0^1d\tau 
e^{-i\tau \omega }\left( \partial _\mu \omega \right) e^{i\tau \omega }. 
\label{PureGauge} 
\end{equation} 
Define $U_\tau =e^{i\tau \omega }$, $\psi _\tau =U_\tau \psi $, and note $({%
\vec x}\times {\vec \partial )}^3=\partial _\varphi $, we get  
\begin{eqnarray} 
\delta L_q^{3\prime } &=&-\int_0^1d\tau \int d^3x\psi _\tau ^{\dagger 
}\left( \partial _\varphi \omega \right) \psi _\tau  \nonumber \\ 
&=&-\int_0^1d\tau \int d^3x\left\{ \partial _\varphi \left( \psi _\tau 
^{\dagger }\omega \psi _\tau \right) -\omega ^l\partial _\varphi \left( \psi 
_\tau ^{\dagger }T^l\psi _\tau \right) \right\}  \nonumber \\ 
&=&-\int_0^1d\tau \int d^3x\varepsilon ^{ij3}\partial _j\left( 
x_i\psi _\tau ^{\dagger }\omega \psi _\tau \right) \nonumber \\
&&-\int_0^1d\tau \int d^3xi\left[ J_{QCD}^3,\psi 
_\tau ^{\dagger }\omega \psi _\tau \right],  \label{deltaLq3} 
\end{eqnarray} 
where in the last step we have noticed that $\omega ^\alpha $ is a $c$%
-function commuting with $J_{QCD}^3$, and $\psi _\tau ^{\dagger }T^\alpha 
\psi _\tau $ is a scalar operator under spatial rotation, therefore $%
\partial _\varphi \left( \psi _\tau ^{\dagger }T^\alpha \psi _\tau \right) 
=-i\left[ J_{QCD}^3,\psi _\tau ^{\dagger }T^\alpha \psi _\tau \right] $, 
which is analogous to the Heisenberg equation of motion $\partial 
_tO=i\left[ H_{QCD},O\right] $. Another critical point in this observation 
is that, as can be seen from the explicitly gauge invariant expression in 
eq.(\ref{decomp2}), the total angular momentum operator ${\vec J}_{QCD}$ is 
invariant under gauge transformations, i.e., it can be equivalently 
expressed in terms of $\psi $, $A_\mu $ or $\psi _\tau $, $A_\mu ^\tau 
\equiv U_\tau ^{-1}A_\mu U_\tau +\frac igU_\tau ^{-1}\partial _\mu U_\tau $. 
 
The first term in the final expression of eq.(\ref{deltaLq3}) is a surface 
term. Note that as $|{\vec x}|\rightarrow \infty $, $\psi _\tau ^{\dagger 
}\psi _\tau $ vanishes faster than $|{\vec x}|^{-3}$,  
therefore this term vanishes 
after integration. Such surface terms will be encountered again and again 
later in this paper, and can all be shown to vanish by similar arguments as 
here, so we will drop them out without further explanations. 
 
Now all we are left with in eq.(\ref{deltaLq3}) is a commutator, as we aimed 
to show. Thus we have proved that both $G_{N{\vec L}_qN^{\dagger 
}}^{p_1sp_2s}$ and $G_{NN^{\dagger }}^{p_1p_2}$ are SU3) gauge independent. 
Then according to our above explanations following eq.(\ref{G:NN}), $%
\left\langle p_1s\left| {\vec L}_q\right| p_2s\right\rangle $ is SU(3) gauge 
independent. 
 
In the case of U(1) gauge transformation, the pure gauge form $U\partial 
_\mu U^{-1}$ is just $i\partial _\mu \omega $, so some of the above algebras 
become simpler. However, since the proton carries electric charge $e$, $%
N_l(x_1)N_{l^{\prime }}^{\dagger }(x_2)$ acquires an extra phase factor $%
e^{-ie\omega (x_{1)}}e^{ie\omega (x_2)}$ after U(1) gauge transformation. 
But this phase factor makes no trouble: the only modification we need is to 
change the choice of weight-functional in eq.(\ref{rho}) to $\rho [\omega 
]=e^{ie\omega (x_{1)}}e^{-ie\omega (x_2)}/{\rm Det}{\cal R}\left[ \omega 
\right] $; and without modifying any of the above conclusions.%
\footnote{%
This technique does not apply, however, if an operator is not SU(3) invariant.
For example, after the gauge transformation in (\ref{transform}), 
$\psi _\alpha (x_1)\psi _\beta (x_2)$ becomes $U_{\alpha \alpha '}(x_1)
\psi _{\alpha '}(x_1)\psi _{\beta '}(x_2)U_{\beta '\beta}(x_2)$. 
In such case we 
are unable to make a simple remedy by changing the choice of $\rho [\omega]$,
and the above proof of gauge invariance {\em fails}. Of course, further 
investigations might show that such gauge dependence cancels in the ratio of 
three-point to two-point Green functions, leaving the studied matrix element
still gauge independent. However, if this is not the case, then on a color 
non-singlet state, the expectation values of $\vec L_q$, and even the 
expectation values of gauge {\em invariant} operators, will be gauge 
{\em non-invariant}; and we get a peculiar and surprising connection between 
gauge invariance and color confinement: {\em gauge invariance requires color
singlet}.
}
Therefore $%
\left\langle p_1s\left| {\vec L}_q\right| p_2s\right\rangle $ is also U(1) 
gauge independent, and we do not have to worry about the physical meaning 
of the labeling of atomic energy levels any more in spite of the gauge 
dependence of the electron angular momentum operator. 
 
We summarize our above analyses as follows: if after an arbitrary gauge 
transformation with parameters $\omega ^{\alpha}(x)$, 
the extra term $\delta O_\omega$ acquired by some {\em gauge 
dependent} operator $O$ can be expressed as commutators with the rotation 
and/or translation generators of the system, then $O$ has {\em gauge 
independent} expectation value on the states with {\em definite} momentum and 
polarization. (Actually what appears in the calculations is $\delta 
O^{\prime }_\omega$, 
the result of applying an ``anti-transformation'' with parameters 
$-\omega ^{\alpha}(x)$ to $\delta O_\omega$, 
as shown in eqs.(\ref{G2}, \ref{deltaLq'}). 
But since applying this ``anti-transformation'' to $(O+\delta O_\omega )$
just gives the original $O$, $\delta O'_\omega$ is just $-\delta O_{-\omega}$;
therefore if $\delta O_\omega$ can be 
expressed as a commutator, so does $\delta O^{\prime }_\omega $.) In the 
remainder of this paper, we will show that all the operators concerning 
nucleon spin, momentum, and mass structures meet this requirement. 
 
First we look at the other two gauge dependent operators concerning spin 
structure: $L_g^3$ and $S_g^3$. Under the above gauge transformation in eq.(%
\ref{transform}), we have (from now on $U_\tau $ is defined as $e^{-i\tau 
\omega }$ instead of $e^{i\tau \omega }$)  
\begin{eqnarray} 
\delta L_g^3 &=&2\int d^3xTr\left\{ UE_iU^{-1}\partial _\varphi \left( 
UA_iU^{-1}+\frac igU\partial _iU^{-1}\right) -E_i\partial _\varphi 
A_i\right\}  \nonumber \\ 
&=&2\int_0^1d\tau \int d^3xTr\left\{ i\left[ E_i^\tau ,A_i^\tau \right] 
\partial _\varphi \omega +\frac 1g\left( \partial _iE_i^\tau -U_\tau \left( 
\partial _iE_i\right) U_\tau ^{-1}\right) \partial _\varphi \omega \right\}  
\nonumber \\ 
&&+\frac 2g\int_0^1d\tau \int d^3xTr\left\{ -\partial _i\left( E_i^\tau 
\partial _\varphi \omega \right) +U_\tau \left( \partial _iE_i\right) U_\tau 
^{-1}\partial _\varphi \omega -\varepsilon ^{ij3}\left( \partial _i\omega 
\right) E_j^\tau \right\}  \nonumber \\ 
&=&2\int_0^1d\tau \int d^3xTr\left\{ \left( \partial _\varphi \omega \right) 
\left( i\left[ E_i^\tau ,A_i^\tau \right] +\frac 1g\partial _iE_i^\tau 
\right) -\frac 1g\varepsilon ^{ij3}\left( \partial _i\omega \right) E_j^\tau 
\right\} ,  \label{deltaLg} 
\end{eqnarray} 
 
Use $\partial _iO=-i\left[ P_{QCD}^i,O\right] $ in addition to $\partial 
_\varphi O=-i\left[ J_{QCD}^3,O\right] $, with $P_{QCD}^i$ the QCD 
translation operator (see eqs.(\ref{Pqcd}, \ref{Pqcd'}) below), all the 
terms can be written as commutators:  
\begin{eqnarray} 
\delta L_g^3 =&&2\int_0^1d\tau \int d^3x\left[ J_{QCD}^3,Tr\left( \omega 
\left[ A_i^\tau ,E_i^\tau \right] +\frac ig\omega \partial _iE_i^\tau 
\right) \right]  \nonumber \\ 
&&-\frac{2i}g\int_0^1d\tau \int d^3x\varepsilon ^{ij3}\left[ 
P_{QCD}^i,Tr\left( \omega E_j^\tau \right) \right] .  \label{Lg:commut} 
\end{eqnarray} 
 
The calculation for $S_g^3$ is much simpler:  
\begin{eqnarray} 
\delta S_g^3 &=&\frac 2g\int_0^1d\tau \int d^3x\varepsilon ^{ij3}Tr\left\{ 
\left( \partial _i\omega \right) E_j^\tau \right\}  \nonumber \\ 
&=&\frac{2i}g\int_0^1d\tau \int d^3x\varepsilon ^{ij3}\left[ 
P_{QCD}^i,Tr\left( \omega E_j^\tau \right) \right] .  \label{deltaSg} 
\end{eqnarray} 
 
Thus according to our above theorem, we come to the conclusion that the 
quark's and gluon's spin and orbital contributions to nucleon spin, which 
are defined as the expectation values of $S_q^3$, $L_q^3$, $S_g^3$, and $%
L_g^3$ on a nucleon state with definite momentum and polarization along the 
third direction, are all {\em separately} gauge independent, even though the 
operators $L_q^3$, $S_g^3$, and $L_g^3$ are not. 
 
It should also be noted that the expectation values of $L_q^3$, $S_g^3$, and  
$L_g^3$ are not separately {\em Lorentz} invariant. In the infinite momentum 
frame of the nucleon (or equivalently, in the light-cone formalism) and the 
light-cone gauge $A^{+}=0$, the first moment of the polarized gluon 
distribution measured in hard QCD processes can be related to the above 
local operator of gluon spin, which in the light-cone coordinates is 
accordingly \cite{Jaffe}  
\begin{eqnarray} 
S_g^{+}&=&\int d^3xM_{S_g}^{+12} \nonumber \\
&=&2\int d^3xTr\left\{ \left( {\vec E}\times {%
\vec A}\right) ^3+A_{\bot }\cdot B_{\bot }\right\} ,  \label{Sg+} 
\end{eqnarray} 
where $M^{\lambda \mu \nu }$ is the angular momentum tensor of QCD and the 
subscript $S_g$ denotes the gluon spin part. The expectation value of $%
S_g^{+}$ on the $\left| ps\right\rangle $ state is obviously also gauge 
invariant, since in our above proving of gauge invariance, the nucleon 
momentum is not constrained. 
 
The discussions for nucleon momentum and mass structures are similar. The 
QCD momentum operator can be written  
\begin{equation} 
{\vec P}_{QCD}=\int d^3x\psi ^{\dagger }\frac 1i{\vec \partial }\psi +2\int 
d^3xTr\left( E_i{\vec \partial }A_i\right) ,  \label{Pqcd} 
\end{equation} 
or in an explicitly gauge invariant form  
\begin{equation} 
{\vec P}_{QCD}=\int d^3x\psi ^{\dagger }\frac 1i{\vec D}\psi +2\int 
d^3xTr\left( {\vec E}\times {\vec B}\right) ,  \label{Pqcd'} 
\end{equation} 
where ${\vec P}_q=\int d^3x\psi ^{\dagger }\frac 1i{\vec \partial }\psi $ is 
the gauge dependent quark momentum operator. By the same techniques as 
above, it is easy to show that $\delta {\vec P}_q$ can be expressed as a 
commutator with ${\vec P}_{QCD}$, and hence the quark contribution to 
nucleon momentum is gauge independent. Since the total nucleon momentum is 
gauge independent, so does the gluon contribution. 
 
The QCD Hamiltonian operator can be written  
\begin{eqnarray} 
H_{QCD} &=& \int d^3x\psi ^{\dagger }\left( {\vec \alpha }\cdot \frac 1i{%
\vec D}+m\beta \right) \psi  \nonumber \\ 
&& +\int d^3xTr\left( {\vec E}^2+{\vec B}^2\right) ,  \label{Hqcd} 
\end{eqnarray} 
where $\int d^3x\psi ^{\dagger }{\vec \alpha }\cdot \frac 1i{\vec D}\psi $ 
is the (gauge invariant) operator for quark kinetic and potential energy, 
and it was shown by Ji that it contributes about one third of the nucleon 
mass \cite{Ji:mass}. On the other hand, the quark kinetic energy operator $%
E_K=\int d^3x\psi ^{\dagger }{\vec \alpha }\cdot \frac 1i{\vec \partial }%
\psi $ is gauge non-invariant, but the extra term it acquires after a gauge 
transformation can also be expressed as a commutator  
\begin{eqnarray} 
\delta E_K &=&-\int_0^1d\tau \int d^3x\psi _\tau ^{\dagger }{\vec \alpha }%
\cdot \left( {\vec \partial }\omega \right) \psi _\tau  \nonumber \\ 
&=&-i\int_0^1d\tau \int d^3x\left[ P_{QCD}^i,\psi ^{\dagger }\alpha 
_i\omega \psi \right] ,  \label{deltaEk} 
\end{eqnarray} 
therefore the quark kinetic energy inside the nucleon is still gauge 
independent. 
 
In summary, we showed that gauge {\em dependent} operators are possible to have 
gauge {\em invariant} matrix elements on some special physical states, and thus 
can have definite physical meanings and experimental significance. This 
includes all the gauge dependent operators encountered in the studies of 
hadron spin, momentum, and mass structures, and also of atomic energy levels.
These operators have gauge independent expectation values on hadron 
or atomic states with definite momentum and polarization. Therefore the 
conventional decomposition of nucleon spin into contributions from the spin 
and orbital angular momentum of quark and gluon is gauge {\em independent}, 
although some of the corresponding operators are not; and similar conclusions 
apply to hadron energy and momentum structures. 
 
We owe special thanks to T. Goldman, X. Ji, W. Lu, G. Sterman, and C.W. Wong 
for fruitful communications. This work is supported in part by the NSF 
(19675018), SEDC and SSTC of China. \\

{\it Note added:} What we obtained in this paper is a non-perturbative proof
of gauge invariance, while gauge invariance to each order of perturbative 
expansion does not necessarily follow, and needs separate studies. Some 
relevant perturbative calculations can be found in ref. \cite{Hoodbhoy}.


\begin{references} 
\bibitem{Balitsky}  I. Balitsky and U. Braun, Phys. Lett. B {\bf 267}, 405 
(1991). 
 
\bibitem{Jaffe}  R.L. Jaffe, Phys. Lett. B {\bf 365}, 359 (1996). 
 
\bibitem{Lee}  T.D. Lee, {\it Particle Physics and Introduction to Field 
Theory} (Harwood Academic Publishers, New York, 1981). 
 
\bibitem{Ji:spin}  X. Ji, Phys. Rev. Lett. {\bf 78}, 610 (1997). 
 
\bibitem{Chen}  X.S. Chen and F. Wang, Commun. Theor. Phys. {\bf 27}, 121 
(1997). 
 
\bibitem{Weinberg}  S. Weinberg, {\it The Quantum Theory of Fields} (Volume 
II), \S 15.5 (Cambridge University Press, New York, 1996). 
 
\bibitem{Huang}  K. Huang, {\it Quarks, Leptons, {\rm \&} gauge fields} 
(2th edition), p.72 (World Scientific, Singapore, 1992); note that there is 
a sign error in Huang's book. 
 
\bibitem{Ji:mass}  X. Ji, Phys. Rev. Lett. {\bf 74}, 1071 (1995). 

\bibitem{Hoodbhoy}  P. Hoodbhoy, X. Ji and W. Lu, hep-ph/9808305.
\end{references}
\end{document}